# Hands-Off Spreadsheets

Colin A. Kerr,
Standard Life plc,
colinkerr.xls@gmail.com
Example files: http://bit.ly/Njl9gG

**Abstract**

*The wealth of functionality in the Excel software package means it can go beyond use as a static evaluator of predefined cell formulae, to be used actively in manipulating and transforming data. Due to human error it's impossible to ensure a process like this is always error free, and frequently the sequence of actions is recorded only in the operator's head. If done regularly by highly paid staff it will be expensive. This paper applies to those spreadsheets which involve significant operator intervention, describes a method that has been used to improve reliability and efficiency, and reports on how it has worked in practice.*

**1 INTRODUCTION**

It's not uncommon to have to 'work' a spreadsheet in order to arrive at the required output. Typically bulk data held elsewhere, perhaps a CSV file, will be loaded into a sheet, followed by filter, sort, adjustment of formulae and other manipulations. In this situation, as they say with cars, the weakest nut is the one behind the wheel [1]. Unreliability is not the only cost, as the operator is often a highly paid professional. The method described here automates and clarifies the process. In brief, Excel reads row oriented data, dropping each row in turn into the same designated cells on a worksheet. These cells are referenced by cell formulae to produce the required outputs which are then written out, and the next record read in. On-sheet parameters allow the relatively few lines of automation VBA code to be largely generic, hence reusable. Sample code is provided to permit a rapid start in trying out the method. (VBA is Visual Basic for Applications, included with Excel precisely for the purpose of automation.)

**1.1 Rationale**

Spreadsheet dangers have been well documented, with excellent books [O'Beirne, 2005] and articles detailing how to reduce them, but recent Financial Services legislation has thrown into question the use of spreadsheets, on account of throughput, speed, and accuracy. Automation is the obvious answer but obtaining a conventional software engineering solution is not always feasible. The first use of this method was to facilitate easy addition of new business rules when it was feared a contract programmer's solution would require frequent maintenance. Other implementations of user configurable business rules in Excel exist [2].

**2. THE CODING OVERHEAD**

The claim is that it's possible to leverage the strengths of Excel spreadsheets with a minimal contribution from VBA code. The significant word is 'minimal', to avoid a variation on the bespoke program solution where the programmer has to understand precisely the business rules before work can even start. Misunderstandings creep in, and the rules often change while coding is underway. The software industry appreciates the importance of user configuration: that is, not 'hard coding', any values that are likely to require later alteration. However going beyond simple items of data, to permit business rules logic, takes a lot of effort and becomes expensive. Fortunately Excel can provide a powerful Business Rules User Interface, and the missing automation component is relatively trivial. Hard coding can also occur in worksheets [Blayney, 2006].

**2.1 Minimal Code**

Coding a routine when the specification is well defined because it omits detailed business rules is easy. The resulting code is generic, hence quick to write, and can be reused, justifying better testing and documentation. In fact since the entire static logic can be expressed using cell formulae, and on-sheet tables can configure the automation, the VBA code can indeed be largely 'dumb' and formulaic. A quite different means of reusing components has been described in Spreadsheet Components For All [Paine, 2008].

**2.2 Documentation**

Configuration tables define the operations which may previously have been known only to a key person. Formulae in cells also deserve explanation. All too often these are long strings combining functions into an expression that is extremely hard to read, check or alter. Such expressions can be formatted more clearly but this requires space. With hundreds of rows in a sheet, the space left to split out formulae is one dimensional, which handicaps attempts to layout each step. For example Nested If statements. Even if spread out across cells to the right it's hard to find an audit-friendly way of portraying the logic. Using just one row for the data means the logic can be spread over two dimensions. The paper on the Lookup Technique to Replace Nested-IF Formulas [2009, Grossman, Ozluk, Gustavson] goes into more detail but it's not clear how this could be applied over many thousands of similar rows using cells populated by applying Auto Fill to the first row's formulae cells. The paper concedes a shortage of cells could be a problem in some cases. (What's the point of documentation? [Pryor, 2006] covers purpose and form of documentation.)

**2.3 On the Worksheet**

As an illustration, consider a CSV file of thousands of sales records from Caesar's Store. It has to be transformed according to a set of business rules, one being represented by conversion of Roman numerals into familiar Arabic numerals. For example MCDLIX should become 1459. There happens to be a reasonable single cell construction to do this [3], but often a business rule requires a long messy string of formulae that's hard to debug or change.

Figure 1 shows how, with plenty of room on the sheet, the steps can be laid out clearly. The same can be done for all the other business rules. It doesn't matter that the results are spread all over the sheet, because we can link to them from the row of cells marshalling the outputs. Thus we have a set of cells that receive data by automation code, any number of calculations or business rules acting on these, and a set of cells which collect the results and will be written out. Figure 2 shows how the input/output might look with one rule.

Proceedings of EuSpRIG 2012 Conference
"The Science of Spreadsheet Risk Management", ISBN: 978-0-9569258-6-2
Copyright © 2012 EuSpRIG (www.eusprig.org) and the Author(s)
Page 2 of 10

| Roman number | | | | MCDLIX |
|---|---|---|---|---|
| 10s | M | C | X | I |
| 5s | | D | L | V |
| Expand 4 | MCCCCLVIIII | MCCCCLVIIII | MCDLVIIII | MCDLIX |
| Expand 9 | MCCCCLVIIII | MCCCCLVIIII | MCDLVIIII | MCDLVIIII |
| count 1s | 1 | 4 | 0 | 4 |
| count 5s | 0 | 0 | 1 | 1 |
| value | 1000 | 400 | 50 | 9 |
| Answer | | | | 1459 |

**Figure 1: Splitting out a complex formula to reveal the individual steps.**

Input Cells

| Id | Item | Colour | Number |
|---|---|---|---|
| 1 | Toga | Purple | MCDLIX |

Calculated Cells

| Id | Roman | Colour | Number |
|---|---|---|---|
| 1 | Toga | Purple | 1459 |

OutputCells
**1,Toga,Purple,1459**

**Figure 2: Example Input/Output areas.**

**2.4 Automation Code**

Figure 2 shows the input and output areas on the worksheet. The 4 input and 1 output cell, have been given range names, InputCells and OutputCells. In the examples which follow the priority is simple VBA in order to demonstrate the method. Thus the output CSV record has been constructed on-sheet, by concatenating the individual output cells, with separating commas, into one called OutputCells. (There is insignificant speed reduction if additional code is used to construct the output record direct from the calculated cells.).

The code below will open input and output files, loop through the input records, placing each one in InputCells, initiate a Calculate, then read OutputCells for output to file. With the filepaths also in named cells (InputFilePath, OutputFilePath), the VBA is generic in the sense that it can now be used to do the same thing for any file, any numbers of input and output fields, and any rules on the sheet. The header line passes through unchanged, which is unsatisfactory, but we are keeping it simple and need a header line later.

```
'Read CSVs, split out columns, paste on sheet, calculate, output CSVs

Sub readCalculateCSV()
Dim inline As String, inputsArr() As String

Open Range("InputFilePath") For Input As #1
Open Range("OutputFilePath") For Output As #2

Line Input #1, inline   'header passes through
Print #2, inline

Do Until EOF(1)
```

```
   Line Input #1, inline
   inputsArr() = Split(inline, ",")          'separate the CSV fields
   Range("InputCells") = inputsArr()         'write the fields to the cells in Range

   Application.Calculate                     'initiate a calculate of cell formulae

   'ouptut cells have been concatenated, with comma separation in cell "OutCells"

   Print #2, Range("OutputCells")
Loop

Close #1, #2
End Sub
```

## 2.5 Sorting Filtering, and Reporting

A report is required from the above CSV showing the total number for each Item/Colour combination but unfortunately there are duplicate entries which must be ignored. We will sort the file so duplicate records are consecutive, and see how to deal with cross row functions. Since Excel has a fast and powerful sort it will be used here. (See appendix for alternatives.) The control table we will use is Figure 3. We don't require sort order or headings as an option here, but include them to illustrate the way a control table allows generic code, so these have been included in the sample code in the appendix. The entire block of cells is named "SortParams".

| Sort In    | G:\Work\Inputs.csv       |
|------------|--------------------------|
| Sort Out   | G:\Work\InputsSorted.csv |
| Headings ? | Ascending/Descending     |
| y          | asc                      |

**Figure 3**

The code for using this was created by recording a macro as the file was opened, Data, Sort performed, and the result saved. The resulting code required some modifications to use the parameter in the control table and is given in the appendix.

In order to identify duplicate records it's necessary to carry forward a copy of the data for comparison with the subsequent record, so a cell "CarryForward" will be named and data copied here immediately after writing OutputCell. (Only the Id is required but we haven't split that out in the code.) A comparison on-sheet, in cell named "Duplicate" will provide the information the code needs to choose whether to ignore the data or output it. Figure 4 shows how this might look. The cell "Duplicate" <u>must now be included in the range OutputCells</u> so that a single Sheet->VBA transfer suffices for both. Our code will not treat Duplicate as part of the output.

|           | OutputCells        |
|-----------|--------------------|
| Duplicate | Data               |
| FALSE     | **1,Toga,Purple,2092** |
|           | CarryForward       |
|           | 1,Toga,White,90    |

**Figure 4 Extension for ignoring duplicates**

A pivot table will form the report, feeding off data written to sheet "RawData", and this requires us to store the output data in an array prior to dumping it to sheet "RawData".

After copying to sheet RawData it's necessary to split text to columns, and finally refresh the pivot table, which will have to be created manually on the first occasion. The code for both these was derived from the macros recorded while stepping through the operations.

The full code is shown in the appendix. Slightly more ambitious code could deal with most criticisms of the above, but as it stands it will process tens of thousands of sales records in seconds, apply complex business rules using on-sheet logic, strip out duplicates, and produce a report.

There's a mental jump involved in appreciating how it's possible to implement cross row functions, as we are so familiar with all the data and intermediate steps being available simultaneously. The above shows this is quite feasible in most circumstances. It's also possible to compare two (sorted) files and report differences by reading from both files onto the sheet, and using a status cell similar to the Duplicate one above, to determine if a record is missing in one or other file. It's time consuming to do this manually.

## 3. FURTHER CONSIDERATIONS

### 3.1 Performance

Excel and VBA are each reasonably fast, but transfer between VBA and the worksheet is relatively slow and has to be minimised. For a process that had been taking tens of minutes or even hours it's unlikely that tens of seconds will be a problem. Nevertheless there are 'tricks' to maximising performance, and though these are well documented on the Internet, some useful ones are mentioned in the appendix. With performance optimised it has been possible to apply complex logic to high volumes of data, with acceptable run times, but vastly improved reliability and ease of maintenance. In one case many hundreds of spreadsheets in tens of workbooks were replaced by an advanced version of the above example, though this was aided by reading from a database rather than CSVs.

### 3.2 Build On Good Practice

Incorporating automation in no way clashes with what is already known about designing robust spreadsheets. Normal good practice can be applied to all the worksheets involved and security considerations are little changed: VBA code will need unprotected cells for writing, but a workbook that was previously manipulated had limited scope for protection. (VBA code can unprotect and protect sheets.) Each layer should be considered separately:

1) business as usual input,
2) control tables (appendix has an example),
3) sheet used to implement the business logic,
4) VBA project,
5) output sheets.

Once automation is employed, there is scope for additional checks, audit, and logging, and standardisation can be increased with reusable modules. There isn't even a one million row limit anymore.

## 3.3 Playing Safe

A useful precaution is to match up the expected column names with the ones found in the CSV. All that is required is a list of the expected headers in a control table which VBA code can compare with the header record. When more ambitious operations which refer to columns are attempted, for example if choice of column(s) was included in the Sort control table above, it becomes tiresome referring to column numbers. A translation table lets code take the strain of this. In the same way that validation should apply to normal user inputs, it's worth including checks in the code which reads control tables, like looking for superfluous spaces or straying beyond the area defined by the range name for the cells. Generic code works here if some conventions or standards are established for the format of tables.

## 3.4 Out Of Bounds

One drawback with allowing users to extend the business rules is that they may not have an appreciation of the effect of scaling up the demands put on the system. Although not a frequent problem it has been noted that when there are no obvious boundary conditions users can become over ambitious in their expectations, and change the control tables without appreciation of the effect on performance.

## 4. USER ACCEPTANCE

When an existing process has been converted the initial reaction has always been delight that a tedious, repetitive and often error prone process has been reduced to seconds. Then follows the worry about what to do if it goes wrong, as no effort is made to train the owners (often very busy professionals) in VBA. Occasionally an adventurous soul will realise it's possible to clone and then alter the controls, in order to automate a similar process, but so far there's been little interest in learning how to use the method to re-engineer additional processes.

A weakness of the system is the feedback when it crashes, or an error situation arises. Invariably these are due to something going wrong outside the system, like the wrong file being supplied as input, but the first reaction is generally to assume there is a fault in the automation code. Error handling with appropriate feedback helps but it's hard to anticipate every type of user misunderstanding or mistake. After a few iterations of blaming the method, confidence emerges and self diagnosis triumphs. Further investigation is required into how to give the feeling of being-in-control to the users from the outset.

To ensure user acceptance, refactoring [O'Beirne, 2010] as part of the conversion has not been attempted until recently. The scale of change that this requires is challenging with even moderately complex workbooks, but early impressions are that by retaining only the unique row or rows in the type of worksheets we are discussing, and using them as the input area, it becomes possible to split up complex cell formulae in small stages, while running frequent regression tests.

A few processes have been built from scratch using the method and have worked well. In particular the ability to run very large volume regression testing has been a major advantage and highlighted the key press interruption 'feature'.(see appendix)

# 5. APPENDIX

## 5.1 Subtotals

The early version of this method attempted to avoid any interaction with sheets once the control information had been read. For subtotals a control table similar to Figure 4 was used to drive a VBA procedure which dumped the final result to sheet. It was relatively hard to write but has proved sufficient for subsequent needs. The example used in this paper suggests Pivot Tables may be a better alternative.

| Subtotals | |
|---|---|
| Subtotal these Amounts | for Column Names |
| Number | Item, Colour |
| Number, Amount | Item |

**Figure 5: Possible format for a control table for subtotals**.

## 5.2 Sorting

The Data/Sort in Excel is fast, and provided the data will fit onto a sheet it's the first choice as used in the example above. Due to the many options in the sort function it's probably not worth attempting a truly general purpose control. It's worth pointing out that although Excel will only permit selection of three columns in a sort, it's possible to sort additional columns sequentially and obtain the same result as if it had been possible to sort all the columns in one go.

The DOS Sort [4] has proved useful. It can be submitted using the VBA Shell command which runs asynchronously: so VBA will continue on regardless, unless it's forced to wait. There is a wait command but it's better is to check the Job Id as shown below. It may be necessary to reformat the data in each record as the options in DOS sort are limited, but as was seen above it is easy to read in, modify, and write out a file, and that could be step one of a multi-step process.

```
Set objshell = CreateObject("WScript.Shell")
jobId = Shell(theApp, vbNormalFocus)
Do
   Application.Wait (Now + TimeValue("0:00:01"))
Loop Until Not objshell.AppActivate(jobId)
```

## 5.3 Speed Hints

The main factor in applying this method is the manner in which data is transferred to and from sheets. If a number of cells are being written onto a sheet, define a range and read them in a single command as shown in the example of processing a CSV. The same goes for reading back. A similar consideration applies to reading control tables. Don't repeatedly fetch parameters but read the entire table at the outset.

Screen updating slows up a process. It can be turned off and on in code.

Application.ScreenUpdating = False
Application.ScreenUpdating = True

It's also possible to switch calculation to manual in the code but do not do this without good reason, as not turning it back on again can lead to wasted time and incorrect results.

### 5.4 Watch Out

Excel is designed to interrupt calculation when keys are pressed [5] but it appears it does not halt VBA code, which can then retrieve data from cells before calculation has completed. This behaviour only came to light when using Alt-Tab to switch applications while a 30 minute, high volume, regression test was running.

CalculationInterruptKey=xlNoKey will avoid this.

Excel will switch the year and month in a date if it results in a valid date, when VBA writes to a cell. It's a huge pitfall because dates like 13/1/2012 will remain the same and light testing may not show up the problem. Formatting the destination cell as text avoids conversion [6].

### 5.5 Going beyond

Such is the flexibility of Excel that it's often a substitute database distributed across many workbooks by way of external cell links. Database skills are a step beyond coding in VBA, but not any more of a step than from cell formulae to VBA, if MS Access is used. VBA code can readily store and retrieve to a database, and the above techniques used on tables of data, rather than CSV files or spreadsheet stores. Many of the harder tasks such as sorting and filtering then become trivial. This suggests a possible route to dealing with the web of interconnections in some workbook systems.

### 5.6 The Code

An input file with Roman numerals can be created using the Excel Roman function.
For the Roman to Arabic business rule – create a lookup table or use [3].
Remember to create sheets "RawData" and "Report". The first run will fail: create the pivot table and run again.

```
Sub ReportFromCSV()
Dim inline As String, inputsArr() As String, outputsArr() As Variant
Dim datArr(60000) As String, recNo As Long

Call SortFile

Open Range("InputFilePath") For Input As #1
Open Range("OutputFilePath") For Output As #2
Application.ScreenUpdating = False
Line Input #1, inline   'header passes through
datArr(recNo) = inline
Print #2, inline
recNo = recNo + 1

Do Until EOF(1)
   Line Input #1, inline
```

```
    Application.StatusBar = recNo
    inputsArr() = Split(inline, ",")       'separate the CSV fields
    Range("InputCells") = inputsArr()      'write the fields to cells

    Application.Calculate         'initiate a calculate of cell formulae

    'ouptut cells have been concatenated, with comma separation in cell "OutCells"
    outputsArr() = Range("OutputCells")
    If outputsArr(1, 1) <> "Skip" Then
       datArr(recNo) = outputsArr(1, 2)
       Range("CarryForward") = outputsArr(1, 2)
       Print #2, outputsArr(1, 2)
       recNo = recNo + 1
    End If
Loop
Close #1, #2

Worksheets("RawData").Activate
Cells.ClearContents
Range(Cells(1, 1), Cells(recNo, 1)) = Application.Transpose(datArr)
Call textToCols
Worksheets("Report").PivotTables("PivotTable1").PivotCache.Refresh
End Sub
```
'The operation "Data, Text to Columns", was recorded to get this macro, renamed as 'textToCols

```
Sub textToCols()
Worksheets("RawData").Activate
Columns("A:A").Select
Selection.TextToColumns Destination:=Range("A1"), DataType:=xlDelimited, _
   TextQualifier:=xlDoubleQuote, ConsecutiveDelimiter:=False, Tab:=False, _
   Semicolon:=False, Comma:=True, Space:=False, Other:=False, FieldInfo _
   :=Array(Array(1, 1), Array(2, 1), Array(3, 1), Array(4, 1), Array(5, 1), Array(6, 1)), _
   TrailingMinusNumbers:=True
End Sub

Sub SortFile()
Dim sortParams() As Variant, headings As String, order As String
sortParams = Range("SortParams")    'This is the entire block of sort params

If sortParams(4, 1) = "n" Then headings = xlNo Else headings = xlYes
If sortParams(4, 2) = "asc" Then order = xlAscending Else order = xlDescending
'Format= splits comma separated data into columns
Workbooks.Open Filename:=sortParams(1, 2)

Range("A:D").Sort Key1:=Columns(1), Order1:=order, Header:=headings, OrderCustom:=1, _
   MatchCase:=False, Orientation:=xlTopToBottom, DataOption1:=xlSortNormal

ActiveWorkbook.Close SaveChanges:=True, Filename:=sortParams(2, 2)

End Sub
```

O'Beirne, P., (2010) Spreadsheet Refactoring Proc. EuSpRIG,
http://arxiv.org/ftp/arxiv/papers/1009/1009.1412.pdf

Paine, J. (2008) Spreadsheet Components For All. Proc. EuSpRIG,
http://arxiv.org/ftp/arxiv/papers/0809/0809.3584.pdf

Pryor, L., (2006) What's the point of documentation? Proc. EuSpRIG,
http://arxiv.org/ftp/arxiv/papers/1011/1011.1021.pdf

[1] Examples of human error at http://www.eusprig.org/stories.htm
Number 083 "…inadvertently, he did not sort the value column."
Number 070 "Cut a percent, paste an apology"

[2] Business Rules in Excel http://www.arulesxl.com/

[3] Roman to Arabic http://www.excelforum.com/excel-2007-help/715477-roman-to-arabic.html

[4] DOS Sort command: http://www.computerhope.com/sorthlp.htm

[5] Interruption of calculation in VBA.
http://social.msdn.microsoft.com/Forums/zh/exceldev/thread/c25e60f3-fbd6-4ced-9b56-9265ecc95577

[6] The Date formatting problem. http://www.fontstuff.com/casebook/casebook02.htm
http://social.msdn.microsoft.com/Forums/en-US/vbbugsubmissionpilot/thread/0d4a1c1d-378b-43c0-97cb-a693e1cb26dc\

Proceedings of EuSpRIG 2012 Conference
"The Science of Spreadsheet Risk Management", ISBN: 978-0-9569258-6-2
Copyright © 2012 EuSpRIG (www.eusprig.org) and the Author(s)
Page 10 of 10